\documentclass{IEEEtran}
\usepackage{cite}
\usepackage{amsmath,amssymb,amsfonts,bm}
\usepackage{graphicx}
\usepackage{textcomp,nicefrac}
\usepackage{algorithm}
\usepackage{algpseudocode}

\def\BibTeX{{\rm B\kern-.05em{\sc i\kern-.025em b}\kern-.08em
T\kern-.1667em\lower.7ex\hbox{E}\kern-.125emX}}
\begin{document}
\title{Steerable Conditional Diffusion for Domain Adaptation in PET Image Reconstruction}
\author{George Webber, Alexander Hammers, Andrew P.~King and Andrew J.~Reader%
\thanks{\scriptsize G.~Webber was funded by the EPSRC CDT in Smart Medical Imaging [EP/S022104/1] and by a GSK Studentship.}%
\thanks{\scriptsize G.~Webber (\texttt{george.webber@kcl.ac.uk}), A.~P.~King and A.~J.~Reader are with the School of Biomedical Engineering and Imaging Sciences, King’s College London, UK.}%
\thanks{\scriptsize A.~Hammers is with the King’s College London \& Guy’s and St Thomas’ PET Centre.}}
\maketitle

\begin{abstract}
Diffusion models have recently enabled state-of-the-art reconstruction of positron emission tomography (PET) images while requiring only image training data. However, domain shift remains a key concern for clinical adoption: priors trained on images from one anatomy, acquisition protocol or pathology may produce artefacts on out-of-distribution data. We propose integrating \emph{steerable conditional diffusion} (SCD) with our previously-introduced \emph{likelihood-scheduled diffusion} (PET-LiSch) framework to improve the alignment of the diffusion model's prior to the target subject. At reconstruction time, for each diffusion step, we use low-rank adaptation (LoRA) to align the diffusion model prior with the target domain on the fly. Experiments on realistic synthetic 2D brain phantoms demonstrate that our approach suppresses hallucinated artefacts under domain shift, i.e. when our diffusion model is trained on perturbed images and tested on normal anatomy, our approach suppresses the hallucinated structure, outperforming both OSEM and diffusion model baselines qualitatively and quantitatively. These results provide a proof-of-concept that steerable priors can mitigate domain shift in diffusion-based PET reconstruction and motivate future evaluation on real data.
\end{abstract}

\vspace{-0.4cm}
\section{Introduction}\label{sec:introduction}
Positron emission tomography (PET) delivers quantitative molecular imaging but necessitates radiation exposure, motivating methods that can maintain image quality at reduced dose. Classical iterative reconstructions such as maximum-likelihood expectation maximization (MLEM) suffer from high variance in the low-count regime, motivating the use of priors learned from high-quality data. Diffusion models (DMs) offer an unsupervised route to powerful image priors and have recently shown strong recent results for PET reconstruction, e.g. the likelihood-scheduled diffusion (PET-LiSch) approach \cite{webber_likelihood-scheduled_2024}. Yet DMs assume training and test data share a common distribution; deviations due to scanner differences, anatomy, or tracer uptake patterns can violate this assumption and risk degrading reconstruction quality by introducing hallucinations.

Steerable conditional diffusion (SCD) \cite{barbano_steerable_2025} is a recently-proposed method for tackling such domain shifts, by adapting the DM prior during reconstruction of a unique (test-time) dataset. We propose combining SCD with PET-LiSch, creating a PET-LiSch-SCD algorithm that enables scan-specific adaptation of the DM prior without retraining.

\section{Theory}\label{sec:theory} 
Measured PET counts $\mathbf{m}\in\mathbb{N}^{D}$ are modeled as independent Poisson variables with mean $\mathbf{q}=A\mathbf{x}+\mathbf{b}$, where $\mathbf{x}$ is the unknown tracer distribution, $A$ is the system matrix, and $\mathbf{b}$ accounts for scatter and randoms. DMs are a class of generative methods that learn to recover clean images by progressively reversing a sequence of artificial noise corruptions.

In PET-LiSch \cite{webber_likelihood-scheduled_2024}, a score network $s_\theta(\mathbf{x},t)$ is trained to remove added Gaussian noise from PET images at multiple noise levels. Training images are normalized to have a mean intensity close to 1, with random intensity scaling applied for robustness. At reconstruction time, we start from a Gaussian noise sample and alternate between two steps: a denoising update based on the learned prior, and a data consistency step that encourages agreement with the measured data.

After each denoising step, Tweedie’s formula gives $\hat{\mathbf{x}}_t$, an estimate of the endpoint image. Because real PET data spans a broader intensity range than the training images, a global scale factor $c$ is estimated from a surrogate MLEM reconstruction (following \cite{singh_score-based_2024}) and applied to $\hat{\mathbf{x}}_t$ before evaluating the Poisson log-likelihood (PLL). Multiple gradient ascent updates are then performed until $c \cdot \hat{\mathbf{x}}_t$ achieves a likelihood value consistent with a precomputed likelihood schedule, which is derived from surrogate MLEM iterations. This scheduling ensures the final image has data fidelity in line with clinical heuristics (i.e., number of MLEM steps) while benefiting from the improved image modeling capability of the DM.

Following data consistency, the updated $\hat{\mathbf{x}}_t$ is mapped back to a noise image iterate using a denoising diffusion implicit model (DDIM) step, ensuring that the reconstruction remains consistent with the diffusion trajectory.

We further extend PET-LiSch to PET-LiSch-SCD by introducing low-rank adaptation (LoRA) \cite{hu_lora_2022} weights $\Delta\theta$ into the score network after training. These additional parameters, initialized to zero, are optimized after each denoising step with Adam such that the PLL of the scaled estimate $c \cdot \hat{\mathbf{x}}_t$ is increased. This allows the prior to adapt on a per-scan basis in a lightweight and memory-efficient manner, without requiring retraining or paired data. See Algorithm~\ref{alg:pet_lisch_lora} for a full summary.

\begin{algorithm}[t]
\caption{PET-LiSch-SCD Reconstruction}
\label{alg:pet_lisch_lora}
\begin{algorithmic}[1]
\Require Measured sinogram $\mathbf{m}$, system matrix $A$, pretrained score network $s_\theta$, number of reverse diffusion steps $N$
\State From initial surrogate MLEM reconstruction, estimate global scale factor $c$ and compute likelihood schedule $\{ \mathcal{L}_1, \dots, \mathcal{L}_N \}$
\State Initialize noisy sample $\mathbf{x}_N \sim \mathcal{N}(0, I)$
\State Initialize LoRA parameters $\Delta\theta \leftarrow 0$
\For{$k=N,\dots,1$}
    \State Denoise: $\mathbf{x}_{k-1} \leftarrow$ reverse diffusion step using $s_{\theta + \Delta\theta}$
    \State Form Tweedie's estimate $\hat{\mathbf{x}}_{t_k}$ from $\mathbf{x}_{k-1}$
    \For{each LoRA optimization step}
        \State Optimize $\Delta\theta$ to increase PLL of $c \cdot \hat{\mathbf{x}}_{t_k}$
        \State Update Tweedie's estimate $\hat{\mathbf{x}}_{t_k}$ using new $s_{\theta + \Delta\theta}$
    \EndFor
    \State Data consistency: adjust $\hat{\mathbf{x}}_{t_k}$ by gradient ascent on PLL to match target likelihood $\mathcal{L}_k$
    \State Reverse Tweedie's estimate to update $\mathbf{x}_{k-1}$ (DDIM-style correction)
\EndFor
\State \Return Final reconstruction $\mathbf{x}_{0}$
\end{algorithmic}
\end{algorithm}
\begin{figure*}[t]
    \vspace{-0.5cm}
    \centering
    \includegraphics[width=\textwidth]{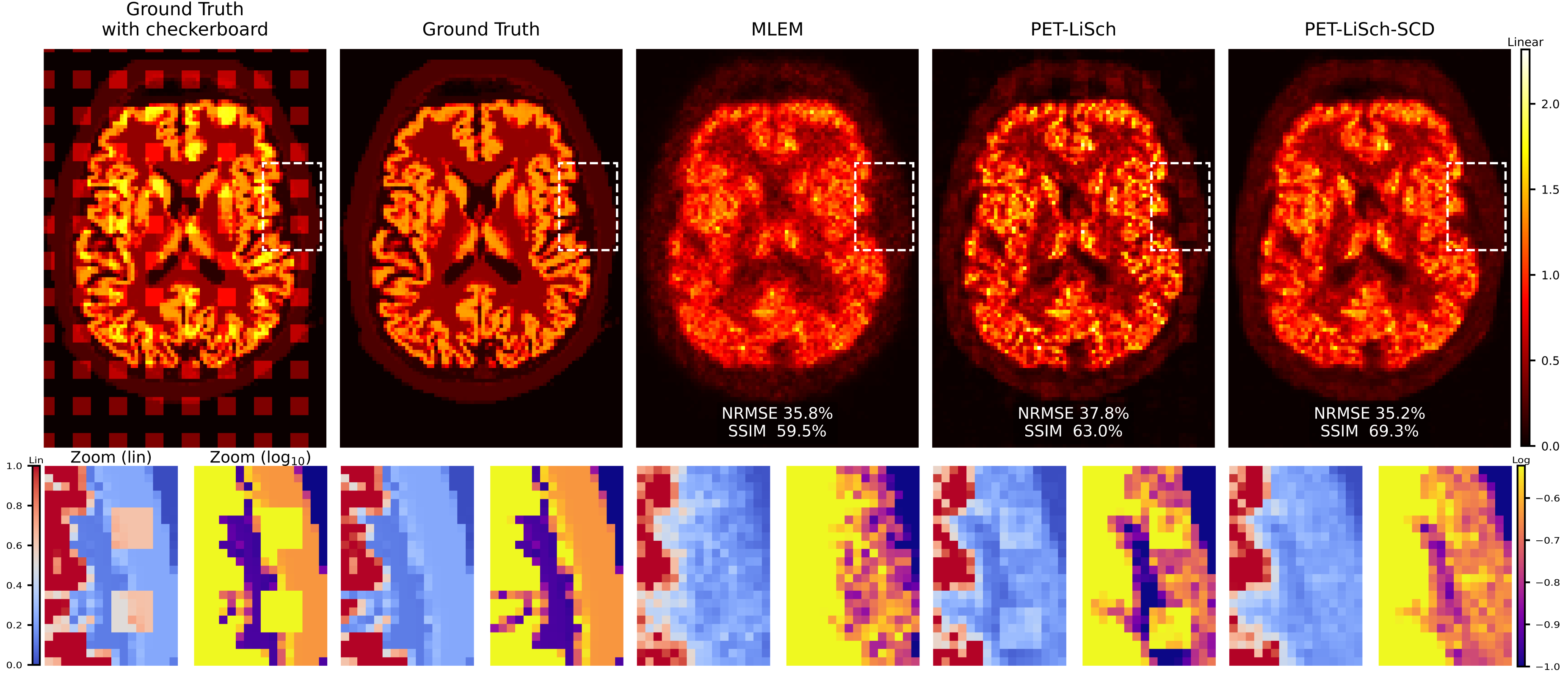}
    \vspace{-0.8cm}
    \caption{Qualitative comparison on low-count reconstructions, with all reconstructions at the same likelihood corresponding to the $40^{\text{th}}$ iteration of MLEM. Column 1: the checkerboard artefact added to training images, shown applied to the ground truth. Column 2: ground truth image. Column 3: image reconstructed by MLEM (40 its). Column 4: image reconstructed with likelihood-scheduled diffusion (likelihood matched to MLEM image). Note the presence of the checkerboard artefact along the right-hand side of the image and elsewhere. Column 5: proposed method PET-LiSch-SCD, without checkerboard artefact.}
    \vspace{-0.2cm}
    \label{fig:recon_grid}
\end{figure*} 
\vspace{-0.3cm}
\section{Experimental setup}\label{sec:experiments}
Thirty central axial slices were extracted from previously acquired T1-weighted MRI scans of 39 healthy volunteers. Grey and white matter were assigned $[^{18}$F]FDG uptake values following \cite{mehranian_model-based_2020} to simulate realistic PET tracer distributions. To introduce a structured domain shift, a fixed $13 \times 13$ checkerboard intensity pattern was overlaid onto the images.

A 2D score network was trained using slices from 38 subjects (split 35:3 for training:validation), following the architecture and noise schedule of \cite{webber_likelihood-scheduled_2024}. Forward projections were simulated using ParallelProj \cite{schramm_parallelproj_2024} through a single-ring Siemens Biograph mMR scanner geometry (span 11, $2\,\text{mm}$ voxels). Scatter and randoms were modeled as a uniform background contributing 30\% of counts, while attenuation correction was calculated from the MRI used to generate the phantom. Poisson noise was applied to a sinogram with $ 6 \times 10^6$ counts, to simulate a low-dose setting.

We compared three reconstruction methods: (i) MLEM with 40 iterations, (ii) PET-LiSch with likelihood matched to the 40-iteration MLEM result, and (iii) the proposed PET-LiSch-SCD (also with matched likelihood). For PET-LiSch-SCD, we applied 8 LoRA optimization steps per diffusion step. LoRA modules were inserted into every convolutional and attention layer, with rank 4 and scale parameter 1. LoRA parameters were optimized using Adam with a learning rate of $1 \times 10^{-4}$.

\vspace{-0.3cm}
\section{Results}\label{sec:results}
Figure~\ref{fig:recon_grid} shows that PET-LiSch reduces noise compared to MLEM but introduces hallucinated checkerboard artefacts from the training domain. PET-LiSch-SCD suppresses these artefacts while maintaining denoising performance and anatomical detail. Quantitatively, only the steered PET-LiSch-SCD method outperformed MLEM at matched likelihood levels, achieving the lowest normalized root mean square error (NRMSE) and highest structural similarity index measure (SSIM) on this test case.

\vspace{-0.3cm}
\section{Summary}\label{sec:summary}
We proposed PET-LiSch-SCD, a steerable diffusion-based method for PET reconstruction that adapts to domain shifts at test time without retraining of the foundation DM. Experiments on synthetic phantoms showed that PET-LiSch-SCD improves robustness compared to both MLEM and unsteered diffusion methods. This approach will need to be extended to 3D datasets and evaluated on real clinical data.


\vspace{-0.3cm}
\begin{thebibliography}{1}
\small
\bibitem{webber_likelihood-scheduled_2024}
G.~Webber \textit{et al.}, ``Likelihood-Scheduled Score-Based Generative Modeling for Fully 3D PET Image Reconstruction,'' \emph{arXiv preprint} arXiv:2412.04339, 2024.

\bibitem{barbano_steerable_2025}
R.~Barbano \textit{et al.}, ``Steerable Conditional Diffusion for Out-of-Distribution Adaptation in Medical Image Reconstruction,'' \emph{IEEE Trans. Med. Imaging}, 2025.

\bibitem{singh_score-based_2024}
I.~R. Singh \textit{et al.}, ``Score-Based Generative Models for PET Image Reconstruction,'' \emph{Machine Learning for Biomedical Imaging: Special Issue on Generative Models}, 2024.

\bibitem{hu_lora_2022}
E.~J. Hu \textit{et al.}, ``LoRA: Low-Rank Adaptation of Large Language Models,'' in \emph{Proc. Int. Conf. Learn. Representations (ICLR)}, 2022.

\bibitem{mehranian_model-based_2020}
A.~Mehranian \& A.~J. Reader, ``Model-Based Deep Learning PET Image Reconstruction Using Forward-Backward Splitting Expectation-Maximization,'' \emph{IEEE Trans. Radiat. Plasma Med. Sci.}, vol.~5, no.~1, 2021.

\bibitem{schramm_parallelproj_2024}
G.~Schramm \& K.~Thielemans, ``Parallelproj: An Open-Source Framework for Fast Calculation of Projections in Tomography,'' \emph{Front. Nucl. Med.}, vol.~3, 2024.

\end{thebibliography}
\end{document}